\documentclass[sigconf,screen]{acmart}

\newcommand{\heading}[1]{\vspace*{1mm}\noindent\textbf{#1.}}
\AtBeginDocument{%
  \providecommand\BibTeX{{%
    \normalfont B\kern-0.5em{\scshape i\kern-0.25em b}\kern-0.8em\TeX}}}

\usepackage{color}
\usepackage{bbm}
\usepackage{multirow}
\usepackage[inline]{enumitem}
\usepackage{graphicx}
\usepackage{subcaption}
\usepackage{sistyle}
\usepackage[ruled]{algorithm2e}
\usepackage{booktabs}
\usepackage{caption}
\SIthousandsep{,}
\usepackage{makecell}
\usepackage{amsmath}
\usepackage[english]{babel}
\usepackage{hyperref}
\usepackage{array} 
\usepackage{graphicx}
\usepackage{algorithmic}
\usepackage{amsmath}
\usepackage{tikz}
\usepackage{wrapfig}

\DeclareMathOperator*{\argmin}{arg\,min}

\makeatletter
\g@addto@macro\normalsize{%
  \abovedisplayskip 0pt plus1pt 
  \belowdisplayskip 0pt plus1pt
  \abovedisplayshortskip  0pt plus1pt%
  \belowdisplayshortskip  0pt plus1pt
}
\makeatother

\copyrightyear{2023}
\acmYear{2023}
\setcopyright{rightsretained}
\acmConference[CIKM '23]{Proceedings of the 32nd ACM International Conference on Information and Knowledge Management}{October 21--25, 2023}{Birmingham, United Kingdom}
\acmBooktitle{Proceedings of the 32nd ACM International Conference on Information and Knowledge Management (CIKM '23), October 21--25, 2023, Birmingham, United Kingdom}
\acmDOI{10.1145/3583780.3614793}
\acmISBN{979-8-4007-0124-5/23/10}

\begin{document}

\title[Black-box Adversarial Attacks against Dense Retrieval Models: A Multi-view Contrastive Learning Method]{Black-box Adversarial Attacks against Dense Retrieval Models:\\ A Multi-view Contrastive Learning Method}

\author{Yu-An Liu}
\orcid{0000-0002-9125-5097}
\author{Ruqing Zhang}
\authornote{Research conducted when the author was at the University of Amsterdam.}
\orcid{0000-0003-4294-2541}
\affiliation{
	\institution{CAS Key Lab of Network Data Science and Technology, ICT, CAS}
	\institution{University of Chinese Academy of Sciences}
	\city{Beijing}
	\country{China}
}
\email{{liuyuan21b,zhangruqing}@ict.ac.cn}

\author{Jiafeng Guo}
\orcid{0000-0002-9509-8674}
\authornote{Jiafeng Guo is the corresponding author.}
\affiliation{
	\institution{CAS Key Lab of Network Data Science and Technology, ICT, CAS}
	\institution{University of Chinese Academy of Sciences}
	\city{Beijing}
	\country{China}
}
\email{guojiafeng@ict.ac.cn}

\author{Maarten de Rijke}
\orcid{0000-0002-1086-0202}
\affiliation{
 \institution{University of Amsterdam}
 \city{Amsterdam}
 \country{The Netherlands}
}
\email{m.derijke@uva.nl}

\author{Wei Chen}
\orcid{0000-0002-7438-5180}
\affiliation{
	\institution{CAS Key Lab of Network Data Science and Technology, ICT, CAS}
	\institution{University of Chinese Academy of Sciences}
 \city{Beijing}
 \country{China}
}
\email{chenwei2022@ict.ac.cn}

\author{Yixing Fan}
\orcid{0000-0003-4317-2702}
\affiliation{
	\institution{CAS Key Lab of Network Data Science and Technology, ICT, CAS}
 \institution{University of Chinese Academy of Sciences}
 \city{Beijing}
 \country{China}
}
\email{fanyixing@ict.ac.cn}

\author{Xueqi Cheng}
\orcid{0000-0002-5201-8195}
\affiliation{
	\institution{CAS Key Lab of Network Data Science and Technology, ICT, CAS}
	\institution{University of Chinese Academy of Sciences}
	\city{Beijing}
	\country{China}
}
\email{cxq@ict.ac.cn}

\renewcommand{\shortauthors}{Yu-An Liu et al.}

\begin{abstract}
Neural ranking models (NRMs) and dense retrieval (DR) models have given rise to substantial improvements in overall retrieval performance.
In addition to their effectiveness, and motivated by the proven lack of robustness of deep learning-based approaches in other areas, there is growing interest in the robustness of deep learning-based approaches to the core retrieval problem.
Adversarial attack methods that have so far been developed mainly focus on attacking NRMs, with very little attention being paid to the robustness of DR models.

In this paper, we introduce the adversarial retrieval attack (AREA) task. 
The AREA task is meant to trick DR models into retrieving a target document that is outside the initial set of candidate documents retrieved by the DR model in response to a query. 
We consider the decision-based black-box adversarial setting, which is realistic in real-world search engines.  
To address the AREA task, we first employ existing adversarial attack methods designed for NRMs. 
We find that the promising results that have previously been reported on attacking NRMs, do not generalize to DR models: these methods underperform a simple term spamming method. 
We attribute the observed lack of generalizability to the interaction-focused architecture of NRMs, which emphasizes fine-grained relevance matching.
DR models follow a different representation-focused architecture that prioritizes coarse-grained representations. 
We propose to formalize attacks on DR models as a contrastive learning problem in a multi-view representation space. 
The core idea is to encourage the consistency between each view representation of the target document and its corresponding viewer via view-wise supervision signals.  
Experimental results demonstrate that the proposed method can significantly outperform existing attack strategies in misleading the DR model with small indiscernible text perturbations. 
\end{abstract}



\maketitle

\vspace{-2mm}
\section{Introduction}

Information retrieval (IR) systems typically employ a multi-stage search pipeline, including the first-stage retrieval and the re-ranking stage~\cite{guo2022semantic}. 
The first-stage retrieval returns an initial set of candidate documents from a large repository, and the re-ranking stage re-ranks those candidates. 
Dense retrieval (DR) models~\cite{zhao2022dense, guo2022semantic} and neural ranking models (NRMs)~\cite{dai2019deeper, xiong2017end} offer substantial performance improvements in the retrieval and re-ranking stage, respectively. 

By modifying normal examples with malicious human-impercep\-tible perturbations, deep learning-based models can be deceived into providing attacker-desired inaccurate predictions~\cite{szegedy2014intriguing}.
DR models and NRMs are prone to inherit the adversarial vulnerability of general neural networks, emphasizing the need for reliable and robust neural IR systems. 
Exploring potential adversarial attacks against neural models in IR is an important step towards this goal: 
Such explorations help identify vulnerabilities, serve as a surrogate to evaluate robustness before real-world deployment, and, consequently, aid in devising appropriate countermeasures.

To date, much attention has been devoted to the design of adversarial attacks against NRMs~\cite{wu2022prada,liu2022order,liu2023topic}. 
Given a neural ranking model, the attack aims to promote a low-ranked target document to higher positions via human-imperceptible perturbations. 
In contrast, little effort has been devoted to investigating how adversarial attacks affect DR models. 
We believe it is important to address this knowledge gap.
Firstly, like NRMs, DR models are increasingly vital in practical IR systems.
Adversarial attacks can expose their weaknesses and provide insights for developing more robust search engines. 
Secondly, within a multi-stage search pipeline, if black-hat search engine optimization practitioners~\cite{gyongyi2005web} cannot ensure a target document successfully passes the first-stage retrieval, they will not have the chance to promote it in rankings in the final ranked list. 

\begin{figure}[t]
    \centering
    \includegraphics[scale=0.88]{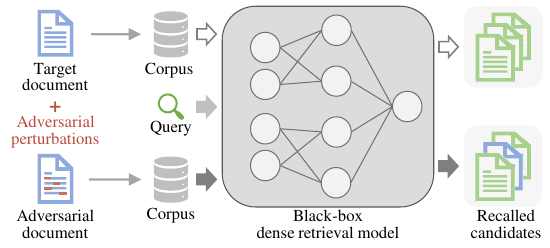}
    \caption{The adversarial retrieval attack (AREA) task.}
    \label{fig:task}
    \vspace{-2mm}
\end{figure}

\heading{Adversarial attacks against DR models}
We are the first to develop adversarial attacks against DR models. 
The first research question is: What is the goal of attacking DR models? 
Based on the adversarial attacks against NRMs and inspired by properties of the first-stage retrieval, we propose to define an attack task, the \emph{adversarial retrieval attack} (AREA) against DR models. 
As shown in Figure~\ref{fig:task}, given a DR model, the AREA task is to retrieve a target document outside the initial set of $K$ candidate documents for a given query, by perturbing the document content in a semantic-preserving way. 
We focus on a practical and challenging decision-based black-box setting, akin to the adversarial attacks against NRMs~\cite{wu2022prada,liu2022order,liu2023topic}, where the adversary can only query the target DR model without direct model information access. 
For consistency with the multi-stage search pipeline in practical IR systems, we simulate a black-box ``retrieval and re-ranking'' pipeline, wherein the target DR model initially narrows down the candidate set to $K$ documents, followed by an NRM determining the final top-$K$ documents ordering. 
In this way, we query the pipeline and assess the final decision to perform attacks in a black-box manner. 

\heading{Using NRM attack methods against DR models}
To address the AREA task, the second research question arises: Do existing attack methods against NRMs perform as well against DR models as against NRMs? 
Our results show that these methods lag behind a simple term spamming attack typically involving query keyword stuffing~\cite{gyongyi2005web}. 
Deep neural networks with interaction-focused architectures are usually employed for NRMs, while less complicated models with representation-focused architectures are adopted in DR models~\cite{guo2020deep,fan2022pre,yates2021pretrained}. 
Specifically, when attacking NRMs, the perturbation update relies on modeling fine-grained interactions between attacked documents and queries. 
In contrast, DR models depend on coarse-grained text representations for effective search in the representation space. 
This distinction renders the existing attacks against NRMs unsuitable for deceiving DR models. 

\heading{Attack models tailored for DR models}
The analysis we have just summarized leads to our third research question: Can we design an effective adversarial attack method tailored for DR models? 
As DR conducts retrieval purely in the representation space~\cite{zhao2022dense,guo2022semantic}, we introduce a \emph{multi-view contrastive learning-based adversarial retrieval attack} (MCARA) to generate adversarial examples. 
Our key idea is to enhance the consistency of semantic representations between the target document and the $K$ retrieved documents in the initial set using view-wise supervision. 
Specifically, after training a surrogate model to demystify the target DR model, we first obtain different viewers to represent documents in the initial set via a clustering technique.  
We produce multi-view representations for the target document through viewers. 
Then, a view-wise contrastive loss is applied to draw each view representation of the target document closer to its corresponding viewer in the semantic space while distancing it from nearest-neighbor documents outside the initial set.  
In this way, the attacker captures informative and discriminative semantic signals via view-level contrastive supervision. 
Finally, following~\cite{wu2022prada}, we use prior-guided gradients of the view-wise contrastive loss to identify the important words in a document, and adopt projected gradient descent~\cite{AleksanderMadry2018TowardsDL} to generate gradient-based adversarial perturbations.

Experiment on two web search benchmark datasets show that MCARA effectively promotes the target document into the candidate set with high attack success and low time cost. 
According to both automatic and human evaluations, MCARA retains target documents semantics and fluency. 
Moreover, the adversarial examples produced by MCARA can deceive the NRM to some extent.   

\vspace{-2mm}
\section{Related Work}
\textbf{Dense retrieval.}
Dense retrieval~\cite{zhao2022dense} conducts first-stage retrieval \cite{guo2022semantic} in the embedding space and has demonstrated several advantages over sparse retrieval~\cite{lin2022proposed}.
It typically employs a dual-encoder architecture to embed queries and documents into low-dimension embeddings~\cite{tonellotto2022lecture}using these similarities as estimated relevance scores~\cite{guo2022semantic}. 
By fine-tuning BERT with in-batch negatives \cite{Karpukhin2020DensePR}, DR models have been shown to outperform BM25~\cite{robertson1994some}.
Subsequently, research has explored various pre-training~\cite{gao2022unsupervised, ma2022contrastive}, and fine-tuning techniques~\cite{Qu2021RocketQAAO,Zhan2021OptimizingDR,Khattab2020ColBERTEA} to enhance DR models, achieving state-of-the-art performance on IR tasks. 
Besides high effectiveness, the robustness of DR models, such as out-of-distribution \cite{thakur2beir,yu2022coco,liu2023robustness} and query variations~\cite{chen2022towards,penha2022evaluating}, has been focused. 
Unlike the work listed above, we focus on the \emph{adversarial} robustness of DR models. 

\heading{Adversarial attacks in IR}
In IR, black-hat search engine optimization (SEO) has been a threat to search systems since the dawn of the world wide web~\cite{gyongyi2005web}.
Black-hat SEO usually aims to increase the exposure of a document owner's pages by maliciously manipulating documents, resulting in a decline in the quality of search results and inundation of irrelevant pages~\cite{castillo2011adversarial}.
Research has shown that neural ranking models (NRMs) inherit the adversarial vulnerabilities of deep neural networks, making them susceptible to small perturbations added to documents~\cite{wu2022neural}.
Research into adversarial attacks against NRMs has been growing, with the goal of promoting a target document in the rankings w.r.t.\ a query via imperceptible perturbations. 
Prior work investigates the vulnerability of NRMs in white-box~\cite{CongzhengSong2020AdversarialSC,wang2022bert} or black-box~\cite{wu2022prada,liu2023topic} scenarios, using word substitution~\cite{wu2022prada} or trigger injection~\cite{liu2022order} as document perturbations. 
Similar to NRMs, DR models are also likely to inherit adversarial vulnerabilities of deep neural networks. 
The adversarial vulnerability of DR models remains under-explored.

\heading{Multi-view document representations} 
A single representation vector may not be able to properly model the fine-grained semantics of a document~\cite{zhao2022dense}.
To tackle this issue, previous work has proposed approaches to explore multiple representations for enhancing the semantic interaction in DR.  
Poly-Encoder~\cite{humeaupoly} learns multi-representations for modeling the semantics of a text according to multi-views. 
\citet{zhang2022multi} introduce multiple viewers to produce multi-view representations to represent documents and enforce them to align with different queries.
In this work, we generate multi-view representations of a target document through viewers. 

\heading{Contrastive learning}
Contrastive learning~\cite{khosla2020supervised} is a branch of self-supervised representation learning in deep learning, which has been widely applied in computer vision~\cite{hadsell2006dimensionality,he2020momentum}, natural language processing~\cite{giorgi2021declutr,shi2019deepchannel}, and social network~\cite{wang2021discover,wang2022micro,liang2023learn,liang2023structure}. 
The key idea is to contrast pairs of semantically similar and dissimilar pairs of data, encouraging the representations of similar pairs to be close and those of dissimilar pairs to be further apart. 
In the context of dense retrieval, some work has adopted contrastive learning in  guiding models to learn more distinguishable representations of documents \cite{xiong2020approximate,ma2022pre}. 
Unlike existing work, we aim to obtain an effective attack signal by pulling each view representation of the target document towards its corresponding viewer, while pushing it away from representations of counter-viewers. 

\vspace{-2mm}
\section{Problem Statement}
Given a query $q$, the aim of first-stage retrieval is to recall a subset of potentially relevant documents from a large corpus $\mathcal{C} = \{d_1,d_2$, \ldots, $d_N\}$ with a total of $N$ documents.
In general, a first-stage retrieval model produces a relevance score $s\left(q,d\right)$ of the query $q$ for each document $d$ in $\mathcal{C}$, and then recalls a set of candidates $R$ by selecting the top-$K$ documents with the highest predicted scores. 
Here, $K$ represents the number of candidates in $R$, which is usually significantly smaller than the corpus size $N$. 
For example, the retrieval model outputs the initial set $R = \{d_{1},d_{2},\ldots,d_{K}\}$ with $K$ candidates if it determines that $s(q,d_1) > s(q,d_{2})> \cdots > s(q,d_{K})> \cdots > s(q,d_N)$. 
In this case, $d_{K}$ possesses the lowest relevance score within $R$. 

\heading{Objective of the adversary} 
The \emph{adversarial retrieval attack} (AREA) task is to fool the DR models into retrieving a target document outside the initial set of $K$ candidates in response to a query appearing in the $K$ initial candidates, by finding an optimized and imperceptible perturbation $p$. 
Formally, given a query $q$ and a target document $d$ out of the initial set, the goal is to construct a valid adversarial example $d^{adv} = d \oplus p$, that can be ranked above the $K$-th position.  
Specifically, $p$ is crafted to conform to the following requirements, 
\begin{equation}
\label{eq:form}
\operatorname{Recall}(q,d \oplus p) \leq K \text{ such that }  p \leq \epsilon,
\end{equation}
where $\operatorname{Recall}(q,d \oplus p)$  denotes the ranking position of the adversarial example recalled by $q$.  
A smaller value of $\operatorname{Recall}$ denotes a higher ranking. 
In this case, the rank position of the original $d$ is larger than $K$. 
$\epsilon$ is the maximum perturbation upper bound of $p$. 
Ideally, the perturbation $p$ should preserve the semantics of document $d$ and be imperceptible to human judges yet misleading to DR models.
In this work, we use the number of word substitutions and the similarity of the substituted words as restrictions.

\heading{Decision-based black-box attacks} 
Since most real-world search engines are black boxes, here, we focus on the decision-based black-box attack setting for the AREA task, where the model parameters are inaccessible to the adversary. 
To align with practical IR systems' multi-stage pipelines, we simulate a retrieval-ranking pipeline by incorporating a representative NRM following the target DR model, refer to as black-box.  
We train a surrogate model~\cite{papernot2016transferability} to imitate the target DR model, by querying the pipeline for the final ranking.   


\vspace{-1mm}
\section{Our Method}
We first analyze the difference between attacking NRMs and DR models, and then introduce our attack method for AREA task. 

\vspace{-2mm}
\subsection{Representation and interaction behavior}\label{4}

To address the AREA task, it is natural to consider existing attack methods designed for NRMs. 
However, as our experimental results of Section~\ref{7.1} show, unlike the success in NRMs, these methods designed for NRMs do not achieve promising performance. 
Below, we investigate the potential reasons from several perspectives. 

\heading{Different model architectures in DR models and NRMs} During first-stage retrieval the aim is to discriminate a small set of candidate documents from (potentially) millions of documents in a \textit{coarse-grained} way~\cite{guo2022semantic}. 
To this end, DR models with their representation-focused architectures (i.e., dual-encoder) are extensively adopted to evaluate relevance based on high-level representations of each input text and to ensure efficiency~\cite{zhao2022dense}.  
In contrast, the re-ranking stage conducts \textit{fine-grained} relevance matching between a query and a small set of candidate documents~\cite{guo2020deep}. 
To this end, NRMs with their interaction-focused architectures (i.e., cross-encoder) are widely used to directly learn from interactions rather than from individual representations and to maintain good system performance~\cite{yates2021pretrained}. 

\heading{Different guidelines when attacking DR models and NRMs} To promote a target document in rankings, attacks on NRMs leverage the interaction signals with attention across the query and the target document tokens.  
The adversary captures the signal of \textit{inner-document representativeness}~\cite{ma2021prop,Qu2021RocketQAAO}, which guides the computation of the update direction for adversarial perturbation. 
In contrast, when attacking DR models, it is important to consider \textit{inter-document representativeness}~\cite{ma2022pre,lu2021less}, since the dual-encoder architecture enables the encoding of queries and documents independently. 
To include the target document in the initial candidate set, the adversary aims to find a minimal perturbation that maximizes the probability of the DR model in distinguishing the target document from millions of documents in the embedding space. 

\smallskip\noindent%
In summary, the variation in model architectures and attack supervision signals pose considerable challenges when attempting to deceive DR models using attacks intended for NRMs. Consequently, it is important to develop attack techniques tailored for DR models. 

\begin{figure*}[t]
    \centering
    \includegraphics[scale=0.74]{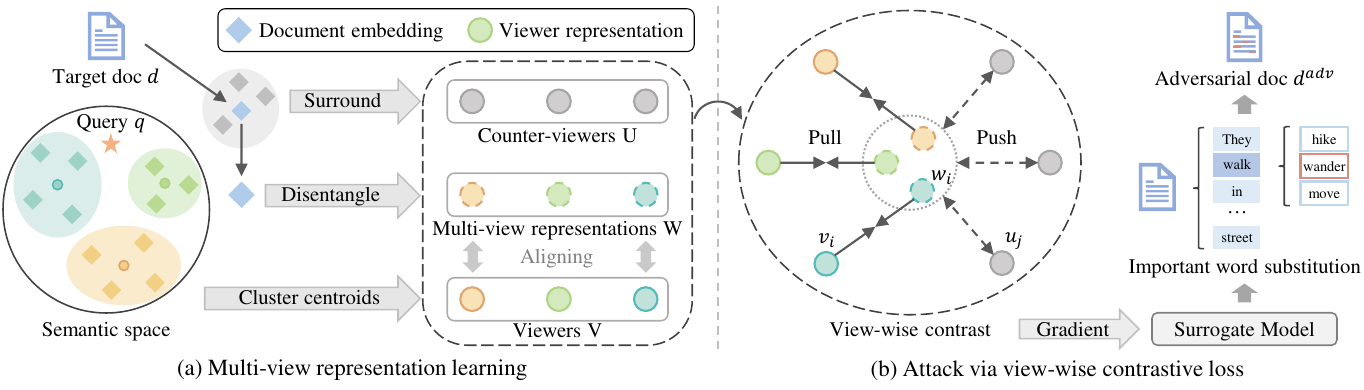}
    \caption{The overall architecture of MCARA. After training the surrogate retrieval model: (a) We learn the multi-view representations of the target document by identifying viewers and counter-viewers. (b) During the attack, a view-wise contrast is used to force each view of the target document close to its corresponding viewer, while away from other counter-viewers.}
    \label{fig:framework}
     \vspace{-3mm}
\end{figure*}

\subsection{Overview of MCARA}

\noindent%
High-quality text representation is the foundation of DR~\cite{ma2022pre}.
We propose to formalize the AREA task as a contrastive learning problem~\cite{khosla2020supervised} in the representation space:   
\begin{enumerate*}[label=(\roman*)]
    \item Push target document $d$ away from other documents outside the initial set; and 
    \item Pull $d$ closer to the candidates inside the initial set. 
\end{enumerate*}
However, contrasting all documents inside and outside the initial set incurs computational overhead and lacks directional control.
In this paper, we introduce representative viewers for the $K$ candidates in the initial set and use the nearest neighbors of the target document in the semantic space as counter-viewers to serve as counterexamples.

Considering the viewers, a simple method to conduct contrastive learning is to directly encourage the representation of the target document and that of each viewer in the semantic space to be closer while keeping counter-viewers away. 
Nevertheless, such simultaneous attraction in multiple directions towards a single document representation could potentially lead to information loss. 
Here, we introduce a novel \emph{multi-view contrastive adversarial retrieval attack} (MCARA). 
The key idea is to disentangle the target document embedding into multi-view representations through viewers, and then enhance the consistency between each view representation and the representation of its corresponding viewer. 
MCARA can be decomposed into three dependent components: 
\begin{enumerate*}[label=(\roman*)]
    \item A \textbf{surrogate model imitation} trains a surrogate retrieval model to prepare for a gradient attack; 
    \item \textbf{Multi-view representation learning} finds the viewers and counter-viewers, and generates multi-view representations for the target document; and
    \item \textbf{Attack via view-wise contrastive loss} generates the embedding space perturbations by calculating the gradients of the surrogate model via contrast.  
\end{enumerate*}
The overall architecture of MCARA is shown in Figure~\ref{fig:framework}. 

\vspace{-2mm}
\subsection{Surrogate model imitation}
To simulate a realistic scenario, we regard the ``retrieval and re-ranking'' pipeline as a unified black-box model, where the retrieval model serves as the target DR model for our attack.  
For each dataset used in this study, we first train the target DR model and then train the NRM based on the retrieved candidates given by the DR model.
We use state-of-the-art models~\cite{gao2022unsupervised,ma2021prop} as the backbone of the DR model and the NRM in the pipeline, respectively.
By sending queries to the black-box pipeline and obtaining the ranked list (given by the NRM), following~\cite{wu2022prada,chen2023towards}, we leverage the relative relevance information among the ranked list~\cite{dehghani2017neural} to construct a synthetic dataset, for training a surrogate retrieval model. 

Given a query $q_c$ from a pre-collected query collection $\mathcal{Q}$ that accesses the black-box pipeline, we get the ranking result $R_a$ of $K$ documents returned by the pipeline.   
We generate pseudo-labels as the ground-truth by treating the first $\ell$ ranked documents $R_a[:\ell]$ as relevant documents $R_a^{+}$.
Generally, training a well-performed DR model needs to combine random negative sampling and hard negative sampling~\cite{xiong2020approximate}. 
Therefore, we treat the other documents $R_a[\ell+1:K]$ as hard negative examples $R_a^{-}$, and the ranked documents of other queries except for $q_c$ in $Q$ are regarded as random negative examples.
We initialize the surrogate retrieval model $\tilde{f}$ using the vanilla BERT.  
The relevance score calculated by the surrogate retrieval model is $\tilde{f}\left(\cdot,\cdot\right)$.
We train $\tilde{f}$ by optimizing a pairwise loss function as the negative log-likelihood of relevant documents: 
\begin{equation}
\begin{split}
\label{eq:surrogate_model_training}
\mbox{}
\hspace*{-1mm}
\mathcal{L} \!=\! - \frac{1}{|\mathcal{Q}|}\!\sum_{q_c \in \mathcal{Q}} \!\!\log \frac{ \tilde{f}(q_c,R_a^{+})}{ \tilde{f}(q_c,R_a^{+}) {+} \tilde{f}(q_c,R_a^{-}) {+} \sum_{R'_c \in \mathcal{R}_{/ \{R_a\}}} \tilde{f}(q_c,R'_c) },
\hspace*{-1mm}
\mbox{}
\end{split}
\end{equation}
where $\mathcal{R}$ denotes the set of query collection's ranking results, and $R'_c$ is the ranking result of other queries. 

\vspace{-1mm}
\subsection{Multi-view representation learning} 

Based on the surrogate model $\tilde{f}$, we first learn multiple viewers from representations of $K$ documents within the initial set returned to a query $q$, and then generate multi-view representations to represent the target document $d$ through the learned viewers. 
In addition to the viewers, we employ a set of counter-viewers from representations of documents outside the initial set to prepare for attacks. 

\heading{Deriving multiple viewers from the initial set of $K$ candidates} 
The key idea is to find several indicative viewers to represent the documents within the initial set and provide guidance for the attack process. 
Here, the viewer is defined as a cluster of documents sharing the same topic. 
We will try other ways of finding viewers in the future. 
Given a query $q$, we first obtain the initial set $R$ of $K$ candidates from the simulated pipeline. 
Then, we use the document embedding generated by $\tilde{f}$ as the representation of each document in $R$.
We apply clustering to the representations of $K$ candidates to obtain $n$ clusters where $n \ll K$, and leverage the representation of each centroid as a topical viewer.  

Specifically, given the $K$ documents in the initial set $R$, we use the K-Means clustering algorithm~\cite{sculley2010web} to find a set $V$ of $n$ viewers, 
\begin{equation}
\begin{split}
\label{eq:kmeans}
V =  \operatorname{Kmeans}\left(n,\varphi\left(\tilde{f},R\right)\right) .
\end{split}
\end{equation}
Here, $\varphi(\tilde{f},R)$ are the embeddings of all $K$ documents in $R$, with respect to the surrogate model $\tilde{f}$.
In this way, we can obtain the representations of $n$ viewers, denoted as $V = \left\{\boldsymbol{v_1},\boldsymbol{v_2},\ldots,\boldsymbol{v_n}\right\}$.

\heading{Generating multi-view representations of the target document through viewers} 
The key idea is to disentangle the view information of the target document aligning to the given viewers, enabling us to effectively extract the specific relevance signal within the candidate set. 
We use a fully-connected layer with activation function ReLU as a multi-view representations generator.
We feed the target document embedding $\boldsymbol{w_d}$ obtained by $\tilde{f}$ and the representations of viewers $V$ into the generator.
To obtain $n$ multi-view representations $W = \left\{\boldsymbol{w_1},\boldsymbol{w_2},\ldots,\boldsymbol{w_n}\right\}$ aligned to viewers, following~\cite{colombo2022learning, cheng2020improving}, we encourage the $\boldsymbol{w_i}$
and its corresponding viewer $\boldsymbol{v_i}$ to be similar while retaining the original information by minimizing the square loss, i.e.,
\begin{equation}
\mathcal{L}_{squ} =  \sum_{i=1}^{n} \left( \Vert \boldsymbol{w_i} - \boldsymbol{v_i} \Vert_2^2 + \Vert \boldsymbol{w_i} - \boldsymbol{w_d} \Vert_2^2 \right),
\end{equation}
where $\boldsymbol{v_i}$ refers to the $i$-st viewer representation and $\boldsymbol{w_i}$ denotes $i$-st the disentangled view representation of target document. 

We maintain the distinction between multi-view representations by maximizing the cosine similarity between them: 
\begin{equation}
\begin{split}
\mathcal{L}_{cos} =  -\sum_{i=1}^{n} \sum_{j=1}^{n} \mathbbm{1}_{[i\neq{j}]} \frac{\boldsymbol{w_i} \cdot \boldsymbol{w_j}}{ \Vert \boldsymbol{w_i} \Vert_2 \Vert \boldsymbol{w_j} \Vert_2}. 
\end{split}
\end{equation}
Combining the two optimization objectives, the multi-view representations $W = \left\{\boldsymbol{w_1},\boldsymbol{w_2},\ldots,\boldsymbol{w_n}\right\}$ of $d$ are calculated by, 
\begin{equation}
\begin{split}
\label{eq:view rep}
W = \argmin  \left(\mathcal{L}_{squ} + \lambda \mathcal{L}_{cos}\right), 
\end{split}
\end{equation}
where $\lambda$ is a trade-off parameter.

\heading{Obtaining multiple counter-viewers from dynamic surrounding documents} 
To enable a contrastive learning based attack, we also propose to find the counter-viewers from documents outside the initial set $R$, pushing the target document away from its original position in the representation space.  
To achieve this goal, we use the dynamic surrounding documents of the target document as counter-viewers for contrast. 
During the attack process, a dynamic surrounding document $d^s$ is the document among the top-$n$ nearest-neighbor to the current perturbed document in the semantic space of the surrogate model $\tilde{f}$. 
We collect the embedding of each dynamic surrounding document:
\begin{equation}
\begin{aligned}
\label{eq: surrounding}
U = \left\{\tilde{f}(d^s) \middle| d^s \in \left\{\kappa\left(\mathcal{C},d,n\right) \right\} \setminus R \right\} ,
\end{aligned}
\end{equation}
where $\kappa(\cdot)$ is a function returning top-$n$ documents closest to the target document $d$ in corpus $\mathcal{C}$ under the semantic space of $\tilde{f}$, and $\tilde{f}(d^s)$ is the embedding of $d^s$.
Finally, we get the representations of $n$ counter-viewers, denoted as $U = \left\{\boldsymbol{u_1},\boldsymbol{u_2},\ldots,\boldsymbol{u_n}\right\}$.

\vspace{-1mm}
\subsection{Attack via view-wise contrastive loss}

Based on the multi-view representations of the target document, viewer representations and counter-viewer representations, we describe how to achieve the attack using a view-wise contrast loss.

\heading{View-wise contrastive loss} The view-wise contrastive loss aims to pull each view representation of the target document close to its corresponding viewer, and push it away from the representations from all counter-viewers. 
Given a query, we aim to find the optimal attack direction for the target document under the semantic space of the surrogate model $\tilde{f}$ with a contrastive loss $\mathcal{L}_{CL}$:
\begin{equation}
\begin{aligned}
\label{eq: contrast}
\mbox{}\hspace*{-1mm}
\mathcal{L}_{CL} \!=\! -\sum_{i=1}^n\!\log \frac{ \operatorname{exp}(\operatorname{sim}(\boldsymbol{w_i},\boldsymbol{v_i})/\tau)}{\operatorname{exp}(\operatorname{sim}(\boldsymbol{w_i},\boldsymbol{v_i})/\tau) + \sum\limits_{\boldsymbol{u_j} \in U} \operatorname{exp}(\operatorname{sim}(\boldsymbol{w_i},\boldsymbol{u_j})/\tau)},
\hspace*{-1mm}\mbox{}
\end{aligned}
\end{equation}
where $\boldsymbol{v_i}$ is a viewer representation in $V$ from Eq.~(\ref{eq:view rep}), $\boldsymbol{w_i}$ is a view representation in $W$ from Eq.~(\ref{eq:kmeans}), $\boldsymbol{u_j}$ is a counter-viewer representation in $U$ from Eq.~(\ref{eq: surrounding}), $n$ is the number of viewers, $\operatorname{sim}(\cdot)$ is the dot-product function, and $\tau$ is the temperature hyperparameter.

\heading{Perturbation word selection} As demonstrated in~\cite{wu2022prada,li-etal-2020-bert-attack}, only some important words in the target document act as influential signals for the final attack performance. Therefore, for each token $h_z$ in the target document, we calculate the gradient magnitude $\boldsymbol{g}^{h_z}$ to the embedding vector of each token in $\tilde{f}$ using $\mathcal{L}_{CL}$, 
\begin{equation}
\begin{aligned}
\label{eq: gradient}
\boldsymbol{g}^{h_z} = \frac{\partial \mathcal{L}_{CL}}{\partial \boldsymbol{e}_{h_z}^{\tilde{f}}},
\end{aligned}
\end{equation}
where the $\mathcal{L}_{CL}$ is the view-wise contrastive loss from Eq.~(\ref{eq: contrast}), and $\boldsymbol{e}^{\tilde{f}}_{h_z}$ is the embedding vector of $h_z$ obtained by $\tilde{f}$.

Then, the word importance $I_{h_z}$ of each token $h_z$ is calculated by $I_{h_z} = \| \boldsymbol{g}^{h_z}_{t} \|^2_2 $.
We only attack the top-$m$ words with the highest importance for each target document $d$, i.e.,
$O = \left\{o_1, o_2, \ldots ,o_m\right\}$.

\heading{Embedding perturbation and synonym substitution}
We adopt the projected gradient descent~\cite{AleksanderMadry2018TowardsDL} to generate gradient-based adversarial perturbations to the embedding space. 
Specifically, for each step $t$ in total iterations $\eta$, we calculate the gradient $\boldsymbol{g}^{d}_{t}$ of $\mathcal{L}_{CL}$ with respect to target document $d$ on embedding space.
After $\eta$ iterations, we obtain the perturbed embeddings $\boldsymbol{e^p}$ of all the important words $O$ in $d$: $\boldsymbol{e^p} = \left\{e_1^p, e_2^p, \ldots ,e_m^p \right\}$ from gradient accumulation.

Then, we substitute the important words with synonyms $S$.
Following~\cite{wu2022prada}, we utilize the embedding similarity of counter-fitted word embeddings~\cite{mrkvsic2016counter} to determine synonyms and employ the same greedy word replacement strategy computed by perturbed important word embeddings $\boldsymbol{e^p}$ and synonym embeddings.
Unlike existing work~\cite{wu2022prada,liu2023topic}, we select words from the documents in the initial candidate set as the pool of potential synonym set $S$.
To further consider semantic and fluency constraints of the perturbed sentence, we use the language model perplexity~\cite{radford2019language} threshold $\rho$ of the sentence containing the replacement word to refine the selection of the synonym set.

\vspace{-1mm}
\section{Experimental Settings}
In this section, we introduce our experimental settings. 

\vspace{-2mm}
\subsection{Datasets}
\heading{Benchmark datasets}
We conduct experiments on two standard dense retrieval benchmark datasets: the \textbf{MS MARCO Document Ranking dataset}~\cite{nguyen2016ms} (MS-MARCO Document) which is a large-scale benchmark dataset for web document retrieval, with about 3.21 million documents, and the \textbf{MS MARCO Passage Ranking dataset}~\cite{nguyen2016ms} (MS-MARCO Passage) which is another large-scale benchmark dataset for web passage retrieval, with about 8.84 million passages. 
The relevant documents to user queries are obtained using Bing, thereby simulating real-world web search scenarios. 

\heading{Target queries and documents}
Following~\cite{wu2022prada,chen2023towards}, for each dataset, we randomly sample 500 Dev queries as target queries for evaluation. 
We adopt three types of target documents outside the initial candidate set, which exhibit different levels of attack difficulty, i.e., \textit{Easy}, \textit{Middle}, \textit{Hard}. 
These documents are sampled from the retrieval results of the target DR model.  
For each target query, we select a total of 30 target documents. 
Beyond the above three separate sets of target documents, for each query, we also incorporate a random sampling of 10 documents from the original pool of 30 target  documents.
These documents are selected to showcase a diverse range of attack difficulties, forming a \textit{Mixture} level. 

\vspace{-2mm}
\subsection{Models}
\heading{Baselines} 
We compare our method with several representative attack methods: 
\begin{enumerate*}[label=(\roman*)]
\item \textbf{Term spamming (TS)}~\cite{gyongyi2005web} randomly selects a starting position in the target document and replaces the subsequent words with terms randomly sampled from the target query. 
\item \textbf{TF-IDF} simply replaces the important words in the target document, which have the highest TF-IDF scores based on the target query, with their synonyms.
\item \textbf{PRADA}~\cite{wu2022prada} is a decision-based black-box  ranking attack method against NRMs via word substitution. 
We use the pairwise hinge loss between the target document and the documents from the initial candidate set of DR models to guide the attack. 
\item \textbf{PAT}~\cite{liu2022order} is an anchor-based ranking attack method against NRMs via trigger generation.    
We use the pairwise loss between the target document and the anchor (top-1 document) of DR models to guide the attack. 
\end{enumerate*} 

\heading{Model variants} 
We implement two variants of MCAR, denoted as
\begin{enumerate*}[label=(\roman*)]
    \item \textbf{MCARA$_{single}$} removes the multi-view representation learning and directly leverages the single document embedding obtained by the surrogate model to contrast with different viewers. 
    \item \textbf{MCARA$_{ind}$} contrasts each viewer representation of the target document with its corresponding viewer independently and then calculates the gradient to find important words accordingly. In this way, we can obtain the intersection of important words found by independent gradient perturbation. 
\end{enumerate*} 

\vspace{-2mm}
\subsection{Implementation details}
For MS-MARCO Document and MS-MARCO Passage, the size $K$ of the initial candidate set is 100 and 1000~\cite{ma2022contrastive,ma2022pre,lu2021less}, respectively. 
To obtain the target documents, for each sampled query in the MS-MARCO Document, the Easy level comprises 10 documents ranked between $[101,200]$, with documents evenly sampled from the range. 
The Middle level includes 10 documents ranked between $[201,1000]$, again with documents evenly sampled from the range. 
The Hard level consists of 10 documents ranked outside the top 1000, with each document randomly selected from those outside top 1000. 
For MS-MARCO Passage, the Easy, Middle, and Hard documents are similarly sampled from the ranking range of $[1000,2000]$, $[2000,10000]$ and outside of the top 10000, respectively.

For the black-box "retrieval and re-ranking" pipeline, we choose a representative DR model called coCondenser~\cite{gao2022unsupervised} as the retriever and also as our target DR model. 
Following~\citet{Zhan2021OptimizingDR}, we fine-tune the pre-trained coCondenser using two-stage hard negatives sampling strategy on the corresponding dataset. 
We choose a representative NRM called PROP~\cite{ma2021prop} as the re-ranker and fine-tune the pre-trained PROP using the relevance labels and the retrieval results given by coCondenser.  
Finally, we use the fine-tuned PROP to re-rank the initial candidate set retrieved by the fine-tuned coCondenser and get the final ranked list for guiding the learning of the surrogate model. 
For surrogate model imitation, we choose vanilla BERT$_{base}$ \cite{devlin2018bert} as the backbone of the surrogate DR model with a dual-encoder architecture.  
For each dataset, we utilize the Eval queries as the pre-collected query collection $\mathcal{Q}$.
We set $\ell$ to 1 due to the average number of relevant documents per query. 

For multi-view representation learning, the number of viewers and counter-viewers $n$ is set to 5 for MS-MARCO Document and 10 for MS-MARCO Passage, respectively.
The trade-off parameter $\lambda$ is 10. 
We train the multi-view representations generator using our target query-document pairs for 1 epoch with a learning rate of 1e-6. 
For attack via view-wise contrastive loss, we set the temperature hyperparameter $\tau$ as 0.1.
The total iterations of attack $\eta$ are 3. 
The perplexity threshold $\rho$ is set to 50 for filtering synonyms that do not fluent in the original text. 
Following \cite{wu2022prada}, the number of substitution words $m$ in MCARA is set to 50 and 20 for the MS-MARCO Document and MS-MARCO Passage, respectively.
For a fair comparison, we maintain the same number of substitutions in all baselines. 
And the trigger length of PAT is set to 10 and 5 for the MS-MARCO Document and MS-MARCO Passage, respectively.

\begin{table*}[t]
\centering
   \caption{Attack performance of MCARA and the baselines; $\ast$ indicates significant improvements over the best baseline ($p \le 0.05$).}
   \renewcommand{\arraystretch}{0.9}
   \setlength\tabcolsep{4.5pt}
  	\begin{tabular}{l l  c c c  c c c  c c c  c c c }
  \toprule
   \multirow{3}{*}{Dataset} & \multirow{3}{*}{Method} & \multicolumn{3}{c}{Easy} & \multicolumn{3}{c}{Middle} & \multicolumn{3}{c}{Hard} & \multicolumn{3}{c}{Mixture} \\ 
      \cmidrule(lr){3-5} \cmidrule(lr){6-8} \cmidrule(lr){9-11} \cmidrule(lr){12-14}
    & & \multicolumn{2}{c}{SRR} & NRS & \multicolumn{2}{c}{SRR} & NRS & \multicolumn{2}{c}{SRR} & NRS & \multicolumn{2}{c}{SRR} & NRS \\
        \cmidrule{3-14} 
       &  & @10 & @100 & @100 & @10 & @100 & @100 & @10 & @100 & @100 & @10 & @100 & @100\\ 
       \midrule
    \multirow{7}{*}{\parbox{1.7cm}{MS-MARCO Document}} 
& TF-IDF  & 16.0 & 40.9 & 32.1 & 11.1 & 28.0 & 23.6 & \phantom{1}4.2 & 14.4 & 13.6 & 10.3 & 28.6 & 23.2 \\
& TS  & 37.8 & 88.1 & 67.5 & 27.2 & 58.0 & 60.3 & 15.1 & 35.8 & 33.5 & 27.1 & 61.1 & 54.6 \\
& PAT  & 26.5 & 70.2 & 52.2 & 13.7 & 36.0 & 32.0 & \phantom{1}7.9 & 27.1 & 26.4 & 16.0 & 43.1 & 36.6 \\
& PRADA  & 28.4 & 74.7 & 56.2 & 18.5 & 43.1 & 37.9 & 11.2 & 33.0 & 33.3 & 10.5 & 50.9 & 42.7 \\
 \cmidrule{2-14}
& MCARA$_{single}$ & 36.8 & 85.1 & 64.8 & 25.6 & 58.6 & 57.9 & 18.3\rlap{$^{\ast}$} & 44.1\rlap{$^{\ast}$} & 41.5\rlap{$^{\ast}$} & 26.9 & 62.9 & 54.7 \\
& MCARA$_{ind}$ & 37.1 & 86.2 & 66.1 & 26.3 & 60.2 & 58.4 & 19.6\rlap{$^{\ast}$} & 45.7\rlap{$^{\ast}$} & 43.9\rlap{$^{\ast}$} & 27.4 & 63.9 & 56.1 \\
& MCARA & \textbf{43.5}\rlap{$^{\ast}$} & \textbf{92.3}\rlap{$^{\ast}$} & \textbf{73.1}\rlap{$^{\ast}$} & \textbf{28.1}\rlap{$^{\ast}$} & \textbf{66.5}\rlap{$^{\ast}$} & \textbf{61.4}\rlap{$^{\ast}$} & \textbf{24.4}\rlap{$^{\ast}$} & \textbf{50.2}\rlap{$^{\ast}$} & \textbf{51.3}\rlap{$^{\ast}$} & \textbf{31.2}\rlap{$^{\ast}$} & \textbf{69.9}\rlap{$^{\ast}$} & \textbf{61.5}\rlap{$^{\ast}$}\\
\midrule

     &  & @100 & @1000 & @1000 & @100 & @1000 & @1000 & @100 & @1000 & @1000 & @100 & @1000 & @1000\\ 
       \midrule
  \multirow{7}{*}{\parbox{1.7cm}{MS-MARCO Passage}}   
& TF-IDF  & 10.2 & 35.2 & 25.1 & \phantom{1}6.4 & 19.8 & 18.3 & \phantom{1}2.1 & 10.5 & 10.3 & 6.1 & 21.6 & 17.8 \\
& TS & 28.6 & 79.0 & 59.1 & 17.2 & 50.8 & 48.7 & \phantom{1}8.4 & 27.6 & 26.9 & 17.8 & 52.0 & 44.4 \\
& PAT & 16.4 & 62.3 & 46.7 & \phantom{1}9.4 & 30.0 & 28.6 & \phantom{1}5.3 & 23.4 & 21.5 & 10.4 & 38.6 & 32.3 \\
& PRADA  & 20.1 & 68.2 & 51.0 & 13.8 & 39.9 & 39.6 & 10.6 & 31.5 & 30.1 & 14.7 & 46.4 & 40.2 \\
 \cmidrule{2-14}
& MCARA$_{single}$ & 24.8 & 74.2 & 56.0 & 16.3 & 48.3 & 46.2 & \phantom{1}9.7 & 31.2 & 29.6 & 16.8 & 51.0 & 43.8 \\
& MCARA$_{ind}$ & 26.5 & 76.3 & 59.4 & 18.8\rlap{$^{\ast}$} & 51.9 & 49.9 & 11.1\rlap{$^{\ast}$} & 35.5\rlap{$^{\ast}$} & 34.9\rlap{$^{\ast}$} & 18.8\rlap{$^{\ast}$} & 54.6\rlap{$^{\ast}$} & 48.0\rlap{$^{\ast}$} \\
& MCARA & \textbf{32.9}\rlap{$^{\ast}$} & \textbf{83.1}\rlap{$^{\ast}$} & \textbf{65.9}\rlap{$^{\ast}$} & \textbf{22.7}\rlap{$^{\ast}$} & \textbf{57.3}\rlap{$^{\ast}$} & \textbf{53.7}\rlap{$^{\ast}$} & \textbf{15.3}\rlap{$^{\ast}$} & \textbf{41.1}\rlap{$^{\ast}$} & \textbf{40.2}\rlap{$^{\ast}$} & \textbf{23.7}\rlap{$^{\ast}$} & \textbf{60.5}\rlap{$^{\ast}$} & \textbf{53.3}\rlap{$^{\ast}$} \\
\bottomrule
    \end{tabular}
    \vspace{-1mm}
   \label{table:Baseline}
\end{table*}

\vspace{-2mm}
\subsection{Evaluation metrics}
\heading{Attack performance} 
We consider two automatic metrics:  
\begin{enumerate*}[label=(\roman*)]
\item Success recall rate (SRR)@$k$ (\%) evaluates the percentage of after-attack documents $d^{adv}$ retrieved into the candidate set $R$ with $k \leq K$ documents. 
Note that the evaluation with $k<K$ is more strict than that with $k = K$.  
\item Normalized Ranking Shifts Rate (NRS)@$K$ (\%) evaluates the relative ranking improvement of after-attacked documents which are successfully recalled into the initial set with $K$ candidates, i.e., $\operatorname{NRS}@K = ({\Pi_d - \Pi_{d^{adv}}})/{\Pi_d} \times 100 \%,$ where $\Pi_d$ and $\Pi_{d^{adv}}$ are the rankings of $d$ and $d^{adv}$ respectively, produced by the target DR model. 
Note that if $d^{adv}$ is not successfully recalled into the initial set of  $K$ candidates, its NRS is set to 0. 
\end{enumerate*}

\heading{Naturalness performance} 
We consider three automatic metrics: 
\begin{enumerate*}[label=(\roman*)]
\item Automatic spamicity detection, which identifies whether target pages are spam. 
Following~\cite{liu2022order}, we adopt the utility-based term spamicity method~\cite{zhou2009osd} to detect the adversarial examples. 
\item Automatic grammar checkers, which compute the average number of errors in the attack documents. 
Specifically, we use two online grammar checkers, i.e., Grammarly \cite{grammarly} and Chegg Writing \cite{chegg}, following the settings in~\cite{liu2022order,liu2023topic}.
\item Language model perplexity (PPL), which measures the fluency using the average perplexity calculated using a pre-trained GPT-2 model~\cite{radford2019language}.
\end{enumerate*}
Furthermore, we leverage the human evaluation, which measures the quality of the attacked documents following the criteria in~\cite{wu2022prada}. 


\vspace{-2mm}
\section{Experimental Results}

In this section, we discuss experimental results, findings and the attack effect between the first-stage retrieval and re-ranking stage discussed in earlier sections
of the paper.

\vspace{-3mm}
\subsection{Are attack methods against NRMs effective against DR models?}\label{7.1}

As shown in Table~\ref{table:Baseline},
\begin{enumerate*}[label=(\roman*)]
\item TF-IDF performs poorly, especially on Hard target documents, indicating that the heuristic method is not able to effectively find the most-vulnerable words that help the target model make judgments. 
\item TS performs moderately well, showing that directly adding spamming with query terms helps improve the relevance between the target document and the query. 
However, spamicity can easily be detected by anti-spamming detection \cite{liu2023topic, liu2022order}.  
We will discuss this further in Section~\ref{naturalness}.
\item When we look at the attack methods tailored for NRMs (PRADA and PAT), PRADA performs better than PAT. 
The reason may be that PRADA considers more documents for the pairwise loss calculation, thus obtaining more comprehensive information about the candidate set. 
\end{enumerate*}

However, both PRADA and PAT perform worse than the simple Term Spamming method on DR models. 
The reason may be that NRMs and DR models have different model architectures and behaviors and thus require different supervision signals to guide the attack process. 
In general, the adversarial attack against DR models is a non-trivial problem for existing attack methods.

\vspace{-2mm}
\subsection{How does MCARA perform on DR models?}
\heading{Overall performance} The performance of MCARA and its variants in the DR attack scenario can be found in Table~\ref{table:Baseline}: 
\begin{enumerate*}[label=(\roman*)]
\item Our MCARA outperforms all the baseline methods significantly, illustrating that it is necessary to attack DR models by capturing the inter-document representativeness in the semantic space. 
Generally, as the difficulty of an attack increases, the performance tends to decrease. 
We will explore more advanced objectives tailored for challenging documents in the future. 
\item In general, attacks on MS-MARCO Document tend to have a higher success rate compared to MS-MARCO Passage. The reason may be that the number of documents addressing the relevant topic is generally smaller than the number of passages extracted from those documents, offering a more focused and concise set of information. 
\item The improvement of MCARA over MCARA$_{single}$ suggests that incorporating multi-view document representations is more beneficial in finding fine-grained semantic information than a single document representation, and thus facilitates better contrasting between the target document and each viewer. 
\item The improvement of MCARA over MCARA$_{ind}$ indicates that optimizing from only one view in the semantic space at a time may lead to disorder in the optimization direction of attacking the target document. 
\end{enumerate*}

\begin{figure}[t]
    \centering
    \begin{subfigure}{0.23\textwidth}
        \centering
        \includegraphics[scale=0.17]{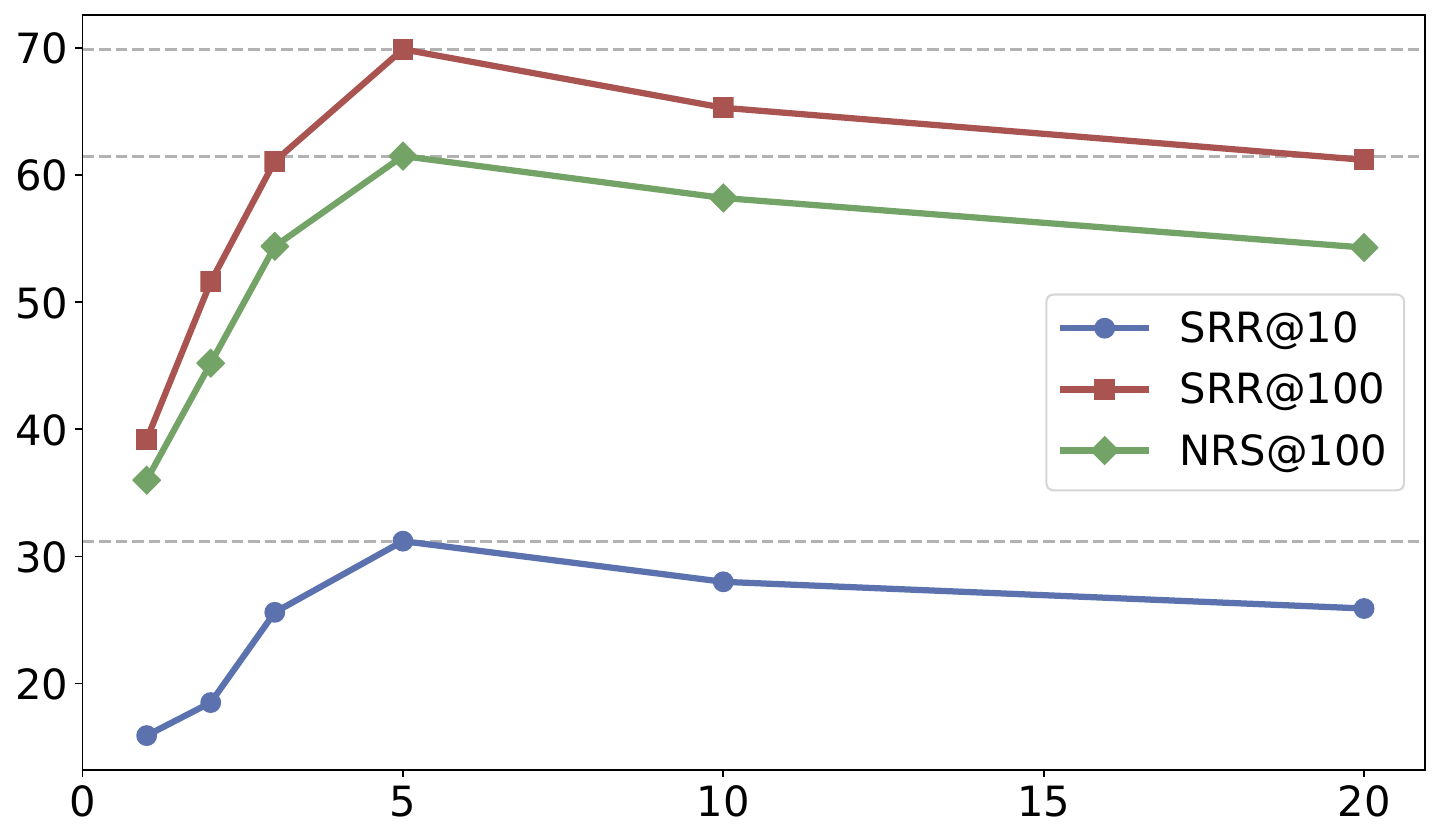}
        \caption*{(a) Number of viewers $\boldsymbol{n}$}
        \label{fig:sub1}
    \end{subfigure}
    \hfill
    \begin{subfigure}{0.23\textwidth}
        \centering
        \includegraphics[scale=0.17]{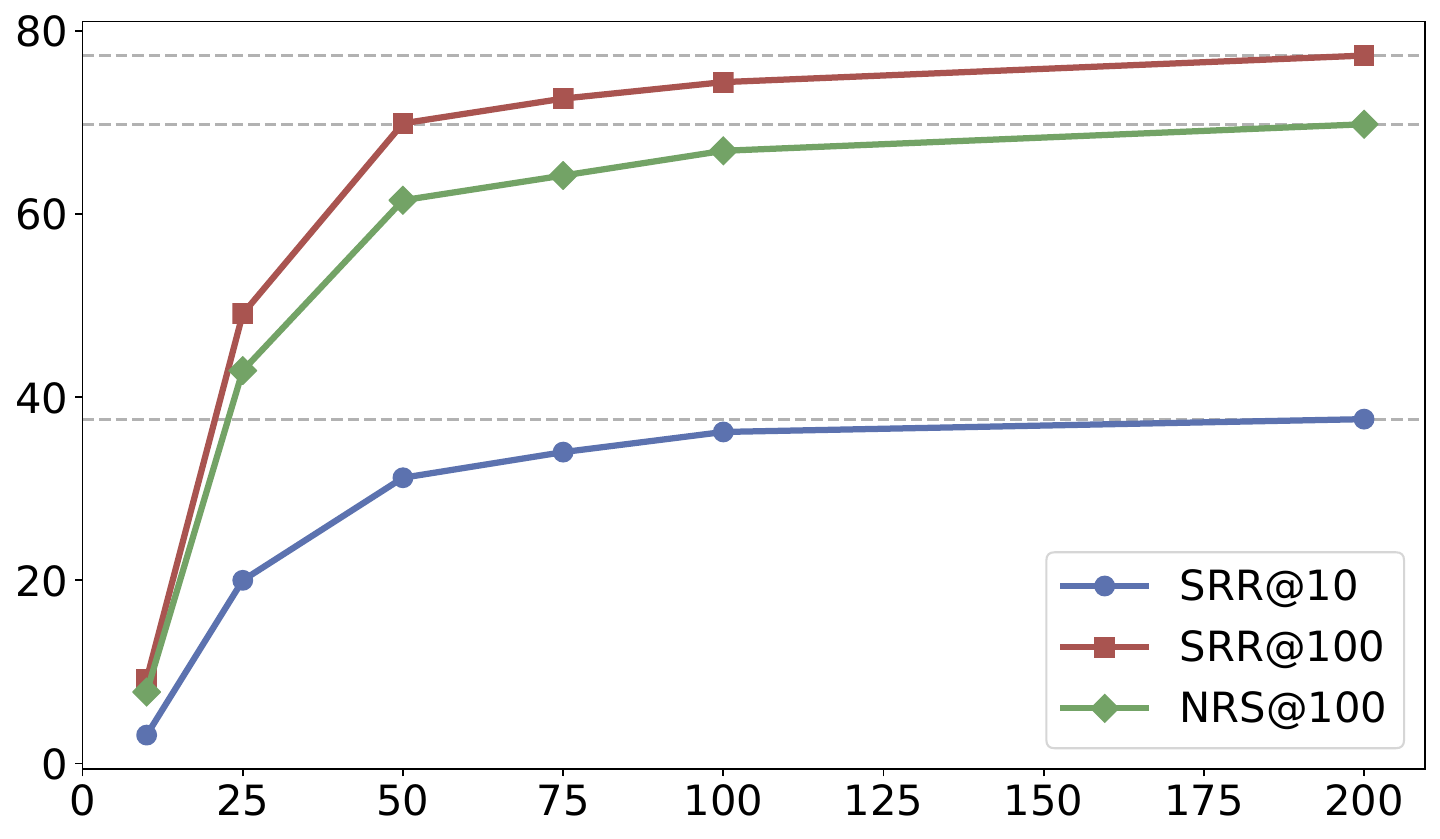}
        \caption*{(b) Perplexity threshold $\boldsymbol{\rho}$}
        \label{fig:sub2}
    \end{subfigure}
    \caption{The impact of the number of viewers (a) and the perplexity threshold (b) on the attack performance of MCARA.}
    \label{fig:both}
    \vspace{-1mm}
\end{figure}

\heading{Impact of the number of viewers} We examine the impact of the important hyperparameter $n$ of MCARA, i.e., the number of viewers, on the attack performance.
The results on the Mixture target documents in MS-MARCO Document are shown in Figure~\ref{fig:both} (a), with similar findings on the other target documents. 
We observe that the performance gets boosted when more representative viewers are incorporated into contrastive learning. 
The reason may be that more viewers can help extract sufficient representative signals for the attack. 
However, the performance gradually decreases when the number of viewers exceeds some threshold. 
Too many viewers increase the risk of making the clusters less representative, even introducing noise which is not good for contrast. 
In the future, we will explore other viewer extraction techniques, such as token embeddings and document-query alignment.

\heading{The impact of the perplexity threshold} We examine the impact of the fluency constrains hyperparameter $\rho$, i.e., the perplexity threshold, on MCARA's performance.
Lower $\rho$ implies tighter substitution fluency constraints in the original sentences. 
The results on the Mixture target documents in MS-MARCO Document are shown in Figure~\ref{fig:both} (b), with similar findings on other target documents. 
Reducing the fluency constraint leads to an improvement in the attack performance, since more disruptive synonyms could be selected. 
However, such a way may lead to the attack easily being detected, as discussed in Section~\ref{naturalness}. 
In the future, it is necessary to investigate more flexible ways to strike a balance between attack performance and the imperceptibility of adversarial perturbations.

\vspace{-2mm}
\subsection{Naturalness of adversarial examples}
\label{naturalness}
We discuss the naturalness of generated adversarial examples with respect to the Mixture-level target documents in MS-MARCO Document, with similar findings on MS-MARCO Passage. 
Here, we remove the fluency constraints of MCARA in the synonym substitution process, denoted as MCARA$_{-FC}$, for comparison. 

\begin{table}[h]
\vspace{-2mm}
\centering
   \caption{The detection rate (\%) via a representative anti-spamming method on the MS-MARCO Document.}
   \renewcommand{\arraystretch}{0.92}
   \setlength\tabcolsep{13pt}
  	\begin{tabular}{l c c c c}
  \toprule
       Threshold & 0.08 & 0.06 & 0.04 & 0.02 \\ 
       \midrule
TS  & 31.4 & 42.2 & 70.3 & 90.1 \\
TF-IDF  & 18.3 & 31.1 & 53.6 & 78.2 \\
PAT   & \phantom{1}8.2 & 13.2 & 24.3 & 46.0 \\
PRADA   & 10.1 & 16.1 & 29.5 & 53.1 \\
MCARA$_{-FC}$ & 11.2 & 17.5 & 31.6 & 55.0 \\
MCARA & \phantom{1}\textbf{6.9} & \textbf{11.6} & \textbf{23.3} & \textbf{44.1} \\
\bottomrule
    \end{tabular}
   \label{table:anti-spamming}
   \vspace{-3mm}
\end{table}

\heading{Automatic spamicity detection} 
Table~\ref{table:anti-spamming} lists the automatic spamicity detection results.  
If the spamicity score of an example is higher than a detection threshold, it is detected as suspected spam content.
We observe that: 
\begin{enumerate*}[label=(\roman*)]
\item As the threshold decreases, the detector intensifies in stringency, leading to an augmented detection rate across all methods.
\item TS can  easily be detected as it integrates numerous repeated query terms into documents. 
\item PAT and PRADA are relatively more undetectable than TS and TF-IDF since they both introduce naturalness constraints.
\item MCARA outperforms the baselines significantly (p-value $< 0.05$), demonstrating the effectiveness of the synonym set derived from the words within candidate documents and the fluency constraints. 
\end{enumerate*}

\begin{table}[h]
\vspace{-2mm}
\centering
   \caption{Online grammar checkers, perplexity, and human evaluation results on the MS-MARCO Document.}
   \renewcommand{\arraystretch}{0.93}
   \setlength\tabcolsep{1pt}
  	\begin{tabular}{l @{} c c c  c c   c c}
  \toprule
       Method & Cheg. & Gram. & PPL & Impercept. & \textit{kappa} & Fluency & \textit{Kendall} \\
       \midrule
Original  & 41 & 63 & 43.1 & 0.85 & 0.56 & 4.32 & 0.61\\
\midrule
TS  & 78 & 120 & 110.6 & 0.06 & 0.42 & 2.12 & 0.79\\
TF-IDF & 76 & 111 & 111.5 & 0.07 & 0.68 & 2.01 & 0.90\\
PAT & 56 & \phantom{1}91 & \phantom{1}62.2 & 0.76 & 0.40 & 3.62 & 0.73\\
PRADA & 67 & 102 & \phantom{1}86.3 & 0.62 & 0.51 & 3.10 & 0.85 \\
MCARA$_{-FC}$ & 72 & 108 & \phantom{1}88.2 & 0.60 & 0.47 & 3.01 & 0.72 \\
MCARA & 51 & \phantom{1}87 & \phantom{1}58.1 & 0.82 & 0.46 & 3.81 & 0.81\\
\bottomrule
    \end{tabular}
   \label{table:human evaluation}
   \vspace{-3mm}
\end{table}

\heading{Automatic grammar checker, PPL, and human evaluation} 
Table~\ref{table:human evaluation} lists the results of the automatic grammar checker, PPL, and human evaluation, including the annotation consistency test results (the \textit{Kappa} value and \textit{Kendall's Tau} coefficient) following \cite{liu2022order,chen2023towards}. 
For human evaluation, we recruit five annotators to annotate 32 randomly sampled Mixture level adversarial examples from each attack method \cite{chen2023towards}. 
Following \cite{wu2022prada}, annotators score the \emph{Fluency} of the mixed examples from 1 to 5; higher scores indicate more fluent examples.  
In terms of \emph{Imperceptibility}, annotators determine whether an example is attacked (labeled as 0) or not (labeled as 1). 
We observe that: 
\begin{enumerate*}[label=(\roman*)]
\item TS performs poorly under all the naturalness evaluations, due to the spamming terms being inserted abruptly in various positions of the document, without considering semantics. 
\item Attack methods with naturalness constraint (i.e., PAT, PRADA) fall short when compared to the original samples, suggesting that making attack examples imperceptible is challenging. 
PRADA underperforms PAT, and a possible reason is that PRADA does not consider the fluency of the sentence with the word replacement.
\item While removing the fluency constraint can enhance the attack performance, adversarial examples generated by MCARA$_{FC}$ raise suspicion.
\item Despite the remaining gap to natural documents, the MCARA with fluency constraint achieves the best naturalness performance among all attack methods. 
\end{enumerate*}

\vspace{-2mm}
\subsection{Can adversarial examples against DR models be promoted in the re-ranking stage?}
Here, we explore the ability of the proposed MCARA to fool NRMs. 
For each dataset, we first obtain the candidate set for each query given by the target DR model, including the successful adversarial examples generated by MCARA.
The size of the candidate set is 100 and 1000 for MS-MARCO Document and MS-MARCO Passage, respectively.
Then, we directly feed these candidate sets to the NRM in the black-box pipeline.
\textit{Avg.rank} measures the average ranking of adversarial examples in the final ranked list and \textit{T50\%} and \textit{T10\%} measure the percentage of adversarial examples entering the top-50\% and top-10\% of the final ranking list. 
As shown in Table~\ref{table: attacks NRMs}, some adversarial examples, as determined by the NRM, are positioned among the high-ranked entries in the final list. 
This suggests that the proposed MCARA method also poses a threat to NRMs.

\begin{table}[h]
   \vspace{-2mm}
\centering
   \caption{The performance of adversarial examples generated by MCARA on attacking against NRMs.}
 \renewcommand{\arraystretch}{0.93}
   \setlength\tabcolsep{4.5pt}
  	\begin{tabular}{l   c c  c c  c c }  \toprule
     \multirow{2}{*}{Tar. Docs} & \multicolumn{3}{c}{MS-MARCO Document} & \multicolumn{3}{c}{MS-MARCO Passage}  \\ \cmidrule(r){2-4} \cmidrule(r){5-7}
      & Avg.rank & T50\% & T10\% & Avg.rank & T50\% & T10\%\\ 
    \midrule
Easy & 68.2 & 34.2 & 5.4 & 736.6 & 28.3 & 3.4  \\
Middle & 82.3 & 15.8 & 2.6 & 874.7 & 11.6 & 1.2  \\
Hard & 91.6 & \phantom{1}3.6 & 0.0 & 958.1 & \phantom{1}2.7 & 0.0 \\
Mixture & 78.6 & 16.7 & 2.1 & 862.1 & 15.0 & 1.7  \\
\bottomrule
    \end{tabular}
   \label{table: attacks NRMs}
   \vspace{-1mm}
\end{table}

\vspace{-2mm}
\subsection{Can adversarial examples against NRMs pass the first-stage retrieval?}

Here, we analyze whether adversarial examples tailored against the NRMs in the re-ranking stage can successfully pass through the first-stage retrieval.   
Following the attack setting in \cite{wu2022prada}, we evenly select nine documents of varying rankings from each ranked list of 200 queries as the target documents for the attack. 
The size of the ranked list is 100 for MS-MARCO Document and 1000 for MS-MARCO Passage, respectively.
We employ PRADA and PAT to attack NRMs and directly feed the adversarial examples into the target DR model.
Here, \textit{Drop} (\%) measures the percentage of adversarial examples, whose rankings are decreased in the first-stage retrieval compared to the original rankings given by the target DR model. 
\textit{NRS} (\%) measures the relative ranking changes of adversarial examples in the first-stage retrieval. 
Here we remove the constraint on the $K$ candidates.
Less than zero indicates an overall decrease in ranking, and greater than zero indicates an overall increase in ranking.
\textit{Lost} (\%) counts the percentage of adversarial examples against NRMs that cannot be recalled in the first-stage retrieval. 

\begin{table}[h]
   \vspace{-2mm}
\centering
   \caption{The performance of adversarial examples against NRMs on attacking against DR models.}
   \setlength\tabcolsep{7.5pt}
  	\begin{tabular}{l  c c c c c c}
  \toprule
  \multirow{2}{*}{Method} & \multicolumn{3}{c}{MS-MARCO Document} & \multicolumn{3}{c}{MS-MARCO Passage} \\ \cmidrule(r){2-4} \cmidrule(r){5-7}
    & Drop & NRS & Lost & Drop & NRS & Lost\\ 
\midrule
PAT  & 30.2 & \phantom{1}7.7 & 22.1 & 28.1 & 10.3 & \phantom{1}6.3 \\
PRADA  & 55.7 & \llap{-}28.4 & 38.6 & 49.6 & \llap{-}20.9 & 10.6 \\
\bottomrule
    \end{tabular}
   \label{table:NRM attack DR}
   \vspace{-2mm}
\end{table}

When applying PRADA for MS-MARCO Document, the rankings of 96.7\% target documents are improved, with an average boost of 40.1\% over the original ranking. 
When these adversarial examples are applied to the target DR model in first-stage retrieval, as shown in Table~\ref{table:NRM attack DR}, 38.6\% of them are not recalled. 
Similar results are observed in MS-MARCO Passage.
In the ``retrieval and re-ranking'' pipeline, only considering the attacks against NRMs could risk target documents failing to be recalled.
This underlines the importance of devising adversarial documents for first-stage retrieval.

\vspace{-2mm}
\subsection{Black-box vs. white-box attack}
In addition to the black-box attack setting, exploring the white-box setting is valuable for gaining a deeper understanding of our method.  
We evaluate the retrieval performance of the surrogate model and the target DR model in the black-box pipeline over all the queries in the Dev sets of the MS-MARCO Document and MS-MARCO Passage, respectively. 
The MRR$@100$ of the target DR model and surrogate model on the MS-MARCO Document are $41.71$ and $38.60$, respectively.
The MRR$@100$ of the whole black-box pipeline is $42.68$.
The MRR$@10$ of the target DR model and surrogate model on the MS-MARCO Passage are $39.82$ and $36.94$, respectively.
The MRR$@10$ of the whole black-box pipeline is $41.46$. 
To simulate a white-box scenario, we designate the target DR model as the surrogate model while keeping other components unchanged in MCARA, denoted as MCARA$_{white}$. 
The result of the Mixture target documents is shown in Table~\ref{table:Ablation}.
The performance of MCARA under the black-box scenario is similar to that under the white-box scenario. 
This suggests that our surrogate model training method for the black-box pipeline can effectively mimic the behavior of the target DR model to execute threatening attacks on it.

\begin{table}[h]
\centering
\vspace{-2mm}
   \caption{Attack performance comparisons of MCARA between the black-box and the white-box attack setting.}
    \renewcommand{\arraystretch}{0.8}
   \setlength\tabcolsep{3.3pt}
  	\begin{tabular}{l  c c c  c c c }
  \toprule
  \multirow{3}{*}{Method} & \multicolumn{3}{c}{MS-MARCO Document} & \multicolumn{3}{c}{MS-MARCO Passage} \\
      \cmidrule(lr){2-4} \cmidrule(lr){5-7}
     & \multicolumn{2}{c}{SRR} & NRS & \multicolumn{2}{c}{SRR} & NRS \\
    \cmidrule(lr){2-3} \cmidrule(lr){4-4}  \cmidrule(lr){5-6} \cmidrule(lr){7-7} 
     & @10 & @100 & @100 & @100 & @1000 & @1000 \\
       \midrule
MCARA & 31.2 & 69.9 & 61.5 & 23.7 & 60.5 & 53.3\\
MCARA$_{white}$ & 33.3 & 71.1 & 63.6 & 24.8 & 61.6 & 54.6\\
\bottomrule
    \end{tabular}
   \label{table:Ablation}
   \vspace{-2mm}
\end{table}

\vspace{-2mm}
\subsection{Time costs of attack methods}
\begin{figure}[t]
    \centering
    \includegraphics[scale=0.226]{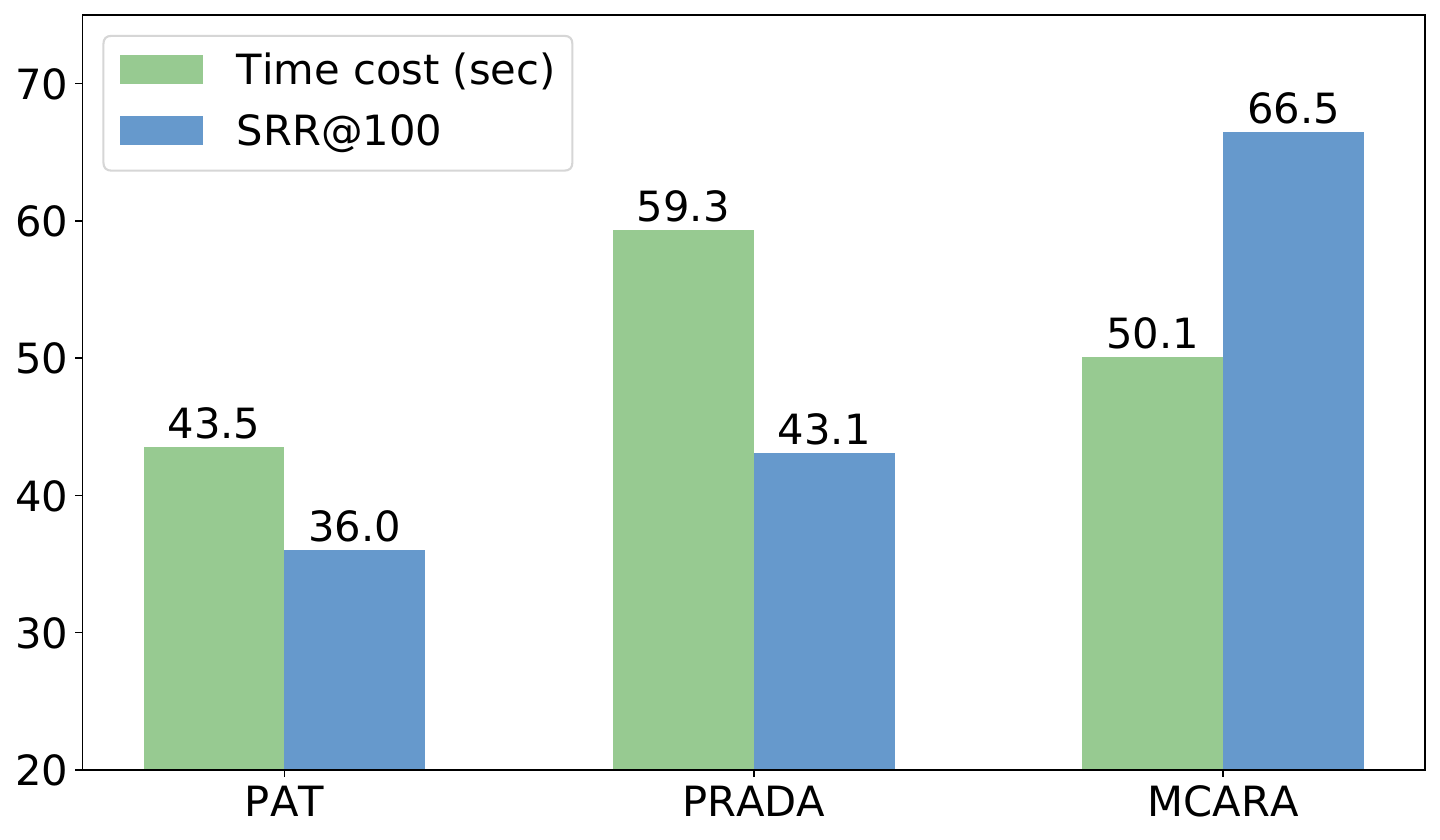}
    \caption{The average time cost of generating an adversarial document and attack performance of different methods.}
    \label{fig:time cost}
\end{figure}

Considering practical application scenarios, an effective attack method should also be efficient, meaning it should find the optimal perturbation with minimal time cost.
Hence, we measure the average time taken by different attack methods to generate an adversarial document using one Tesla V100 GPU.
The results on the Mixture target documents in MS-MARCO Document are shown in Figure~\ref{fig:time cost}, with similar findings on the other target documents.
PAT, relying solely on Anchor document information to optimize perturbations, has lower time costs but poorer attack performance due to insufficient information. 
In contrast, PRADA takes longer due to comparisons with more documents, while MCARA reduces the time overhead while achieving excellent attack results through efficient view-wise supervision.

\vspace{-2mm}
\section{Conclusion}
In this paper, we proposed the AREA task against DR models, demonstrating that by adding small indiscernible perturbations, the adversarial examples can fool the DR models and pass them into the initial retrieval results. 
We developed a novel attack method MCARA, which utilizes view-wise supervision to capture the inner-document representativeness information in DR models for an effective attack. 
The proposed methodology and experimental results reveal the potential risk and vulnerabilities of DR models. 

In future work, it is important to focus on the practical usage of adversarial attacks, specifically for sophisticated real-world search engines that operate with pipelined and ensemble approaches and dynamic corpora. 
A promising direction would involve designing a general unified attack method that can cater to different DR models and NRMs across multiple corpora and modalities. 
Besides, developing effective detection and defense mechanisms against such attacks is crucial for ensuring robustness in IR systems. 

\vspace{-1mm}
\begin{acks}
This work was funded by the National Natural Science Foundation of China (NSFC) under Grants No. 62006218 and 61902381, the China Scholarship Council under Grants No. 202104910234, the Youth Innovation Promotion Association CAS under Grants No. 2021100, the project under Grants No. JCKY2022130C039 and 2021QY1701, the CAS Project for Young Scientists in Basic Research under Grant No. YSBR-034, the Innovation Project of ICT CAS under Grants No. E261090, and the Lenovo-CAS Joint Lab Youth Scientist Project.
This work was also (partially) funded by the Hybrid Intelligence Center, a 10-year program funded by the Dutch Ministry of Education, Culture and Science through the Netherlands Organisation for Scientific Research, \url{https://hybrid-intelligence-centre.nl}, and project LESSEN with project number NWA.1389.20.183 of the research program NWA ORC 2020/21, which is (partly) financed by the Dutch Research Council (NWO).
All content represents the opinion of the authors, which is not necessarily shared or endorsed by their respective employers and/or sponsors.
\end{acks}

\clearpage
\bibliographystyle{ACM-Reference-Format}
\balance
\bibliography{references}

\end{document}